\magnification=\magstep1
\tolerance=10000
\baselineskip=18pt

\centerline{\bf
Efficient quantum computing on low temperature spin ensembles.
}

\vskip 5pt

\centerline{
Stewart E. Barnes
}
\vskip 5pt

\centerline{
Physics Department, University of Miami, Coral Gables, 
FL33124.
}

\vskip 5pt
{\bf 
A new scheme is proposed which will permit electron spin resonance 
pulse techniques to be used to realize a quantum computer with a 100 
qbits, or more.  The computation is performed on effective pure states 
which correspond to off-diagonal blocks of the density matrix.  
Described is a scheme which very efficiently performs the preparation 
stage and which permits ``pseudo-projective measurement'' to be made 
on the output.  With such measurements all members of the ensemble 
remain coherent. }

\vskip 15pt

The demonstration by Shor (1) that a quantum computer can factor large 
numbers in a time which goes as a polynomial rather than the 
exponential of the number of digits has electrified interest in 
quantum computers.  

In a quantum calculation an initial state $|\overline 0> =|0\ldots 0>$ 
is acted upon by a unitary transformation $U$ which reflects the 
calculation and which gives a result $|R> = U |\overline 0>$.  In 
order that the calculation make sense $|\overline 0>$ must be a 
perfectly specified state and it must be possible to determine $|R> $.  
The designer of a quantum computer is immediately faced by three 
non-trivial problems.  First, since $|\overline 0>$ is some particular 
state, the entropy is zero and the system must be cooled to absolute 
zero.  Second, the fundamental tenants of quantum mechanics insist a 
given quantum state $|R> $ cannot be determined accurately by a single 
measurement.  Repeated calculation cycles are necessary 
even before concerns of the signal to noise are addressed.  Third, it 
is necessary to read the individual qbits.

Cirac and Zoller (2) have suggested a physical realization of a quantum 
computer based on linear ion traps, however even supporters (3) of such 
devices as a proof-of-principle for quantum computing concede that 
these will never make a useful computer.  Ingenious schemes for 
performing computations using standard liquid {\it room temperature\/} 
nuclear magnetic resonance (NMR) techniques has been introduced by 
Gershenfeld and Chuang (4) and Cory et al (5).  Again there is a serious 
scaling problem and it is difficult to imagine a  computer with much 
more than 10 qbits.  As pointed out by Warren (6) in a ensemble   of 
$10^{23}$ computing molecules at room temperature there is essentially 
zero probability of finding a single example of any state designated 
to be $|\overline 0>$ for a 100 qbit register.

An efficient $\sim 100$ qbit ensemble spin computer requires an almost 
complete polarization of the spins and almost inevitably low 
temperatures, electronic spins, and the solid state.  Commercial pulse 
electron spin resonance (ESR) spectrometers operating at $\sim 100$ 
GHz are available.  For spins with $g=2$ this corresponds to a field 
of 3T. At a temperature of 1K, $e^{\hbar \omega_{0}/k_{B}T} \sim 
10^{-4}$ so for a 100 qbit computer a large fraction of the computing 
molecules will in the ground state designated to be $|\overline 0>$.  
A basic figure of merit $Q = \omega_{0}\tau$ is the ratio of 
$\omega_{0}$ the transition frequency ($1/\omega_{0}$ is the shortest 
switching time) and $\tau$ the coherence time.  As a general rule, 
light elements have the longest $\tau$, e.g., Li (7) or organic free 
(8) radicals have $g$-factors very close to $2.00$ and very long 
$T_{1}\sim 10^{-3}$s times, implying Q values as large as $10^{9}$ in 
the solid state.  For comparison DiVincenzo (9) gives $Q=10^{7}$ for 
NMR and $Q=10^{13}$ for ion traps.  However these latter figures are 
misleading since the implementation of quantum gates requires 
individual qbits to be addressed.  Both linear traps and liquid NMR 
are {\it very\/} slow because $\Delta \omega_{0}$ the {\it differences\/} in 
frequencies are very small.  In order to avoid this difficulty,
it is envisaged here that each qbit comprises {\it two\/} spins.  
Rare earth, i.e, 4f ions such as Dy and Er often have $g \sim 6-7$ but 
somewhat shorter $T_{1}$ times.  It is imagined that, e.g., a free 
radical plus 4f ion bi-spin unit would be a basic building block with 
the 4f ion always in its very long lived ground state.  The intra-unit 
exchange interaction between this 4f moment and the computing
spin via differing bridge molecules would provide large 
shifts from $g=2$.  The inter-unit exchange provides the necessary 
interactions for qbit manipulation.  The technology envisaged is that 
of molecular magnetism which has been much developed (10) over the 
past decade or so.

The detailed realization of such a machine is not the principal 
concern here.  The questions asked assume that such physical systems 
can be engineered.  Consider the following structure for a register of 
a quantum computer: There is a alternating chain of $N$ computing 
spins which have distinct resonance frequencies $\omega_{n}$ even in 
the absence of interactions.  To simplify the translation to the 
computational problem, for each computing spin the notation 
$|\uparrow> = |1>$ and $|\downarrow> = |0>$ will be adopted and it 
will be assumed that $|0>$ corresponds to the ground state.  Due to 
interactions, each spin sees its two neighbours, so that each 
resonance at $\omega_{n}$ is split into four addressable (well 
separated) lines with frequencies $\omega^{00}_{n}$, 
$\omega^{01}_{n}$, $\omega^{10}_{n}$ and $\omega^{11}_{n}$ as the {\it 
neighbours\/} are $00$, $01$, $10$ and $11$ respectively.  Here $10$ 
means that the spin $n-1$ is $1$ while the $n+1$ spin is $0$.  Because 
of the alternation of the interaction $\omega^{01}_{n} \ne 
\omega^{10}_{n}$.  A $\pi$ pulse at any one of these frequencies amounts 
to the execution of a generalized Toffoli gate, i.e., such a pulse, 
e.g., at the frequency $\omega^{10}_{n}$ will flip the $n$th spin if, and only 
if, its neighbors are ``1'' and ``0''.  This gate along with arbitrary spin 
rotations can be effected by standard pulse resonance technique and 
permit all quantum gates to be constructed.  This has been adequately 
dealt with by others (4,5,11,12) and will not be described here.

Even with the considerable advantage of low temperature electronic 
spin systems there remain two very important problems in using finite 
temperature ensembles for quantum computing.  First, even if they are 
relatively few, it is still necessary to eliminate the signals from 
molecules not in the ground state $|\overline 0>$.  The result $|R> = 
U |\overline 0>$ will be a linear superposition of many states and the 
weight of a given state in $|R>$ will very often be much smaller than 
the weight of unwanted signals from other members of the thermodynamic 
ensemble.  Necessary is a preparation (or final averaging) procedure 
in order that the observable signal originates only from the 
computation pure state $|\overline 0>$.  Most of the procedures 
suggested to date (4,5,12,13) are for small high temperature ensembles.  
They typically become exponentially long for large registers negating 
the advantage of performing a quantum computation.  Potentially 
non-exponential schemes {\it do\/} exist (13) although no specific 
implementation has been suggested for a low temperature ensemble.  The 
principle of these 
earlier techniques is quite different from that proposed here.  The 
usual idea is to reshape the {\it entire\/} density matrix to be an 
effective pure state of the form $({\bf 1}/N) + p|\overline 0>< 
\overline 0|$, while here the aim is to create off-diagonal blocks 
$|\sigma> |\overline 0>< \overline 0|<- \sigma|$, $\sigma = 0,1$ which 
correspond to an ESR signature from a ``signal'' spin reflected by $\sigma$.  
The second problem has to do with projective measurements.  Quantum 
computations typically involve measuring the state of $|R>$ and, after 
the principles of quantum mechanics, this involves a random projection. If 
such measurements {\it were\/} possible using  NMR, or ESR, pulse 
techniques this would leave different members of the 
ensemble in randomly different states. A remedy for this difficulty has been 
suggested (4), however it remains the case that such perfectly projective 
measurements do not exist in the repertoire of NMR, or ESR, pulse 
techniques, although a clever such scheme {\it does\/} exist for ion 
traps (14). It is always possible to imagine deterministic equivalent of a 
given quantum algorithm although no explicit such scheme has yet to be 
presented of the factoring problem.
 Here is proposed a general method by which to perform ``pseudo-projective'' 
measurements and which permit the existing algorithms to used directly.

The basic idea of the new {\it preparation\/} scheme is to perform a simple 
quantum computation on all the states of the ensemble which permits 
the free induction ESR resonance signal to be reduced to that coming 
from the state $|\overline 0>$.  A signal spin, the ESR of which is 
monitored, is added as the first spin of a $N+1$ spin register, thus 
if $|A>$ is an arbitrary ($N$ spin) state, the state of the whole 
register is $|\sigma>|A>$, $\sigma=0,1$.  The equilibrium ensemble 
gives no (free induction decay) ESR signal.  A signal corresponding 
the transitions $|\sigma> \Leftrightarrow |-\sigma>$, for states of 
the form $|\sigma>|0A>$, is turned ``on'' by $\pi/2$ pulse at 
$\omega^{0}_{1}$.  The standard ``spin echo'' technique comprises a 
sequence of gradient-$\pi$-gradient pulses (15).  A field gradient 
applied to the sample causes the spins on different molecules to 
precess at different rates and this destroys the ESR signal, however 
the signal is recovered following a time conjugating $\pi$ pulse if an 
identical gradient pulse is applied to re-focus the spins.  The 
principle of the preparation stage is, between gradient pulses, to 
perform a computation on all the states of the ensemble which time 
conjugates the signal spin if, and only if, the state $|A> = 
|\overline 0>$.  Conceptually the simplest scheme would be to consider 
the signal spin as a separate single bit register $s_{1}$ and perform 
a quantum calculation equivalent to the pseudo code

{\tt If $A =\overline 0$, $s_{1} = s_{1} + 1$ (mod 2)}

The result of a computation is $|\sigma>|R> = |\sigma>U |\overline 0>$ 
where $U$ is the relevant unitary transformation which has no effect 
on the signal spin.  Projective {\it measurements\/} are replaced by what 
might be called ``pseudo-projective'' {\it reversible\/} equivalents.  
The principle (again the practice is somewhat different) is similar to 
the preparation stage.  The presence of a state $|\sigma>|S_{0}> 
\equiv |\sigma> |\sigma_{2}, \sigma_{3} \ldots \sigma_{N} >$ is 
detected by a signal spin ESR by performing the equivalent of 
a calculation on all states $|\sigma>|A>$, between two gradient 
pulses, which reverses the signal spin if, and only if, $A = S_{0}$, 
i.e., the pseudo code is:

{\tt If $A = S_{0}$, $s_{1} = s_{1} + 1$ (mod 2)}

The same pulse sequences in the opposite order return the system to 
the state before the interrogation process and hence this reading 
process is reversible.  Usually the result cannot be determined by a 
single such measurement and measuring every amplitude takes an 
exponentially long time.  Described here are routines which take, at 
the most, $\sim N^{2}$ pulses to read the result for either the Shor or 
Grover (16) algorithms.

While the advantage of a reversible ``read'' process are considerable, 
there is also a major potential disadvantage over projective 
measurements.  After a {\it perfect\/} projective measurement (if such 
a thing really exists), of, e.g., spin orientations corresponding to 
the set of operators $ \hat \sigma^{z}_{n}$, $n=1,N+1$, the system 
finds itself in an eigenstate $\prod_{n=1}^{{N+1}}|\sigma^{z}_{n}>$ of 
these operators.  The weight of this state in original wavefunction is 
reflected not in the strength of the associated ``signal'' but rather 
by the probability of finding the particular signature corresponding 
to the set $\{\sigma^{z}_{n}\}$.  In a straightforward {\it 
reversible\/} measurement of a quantum state the strength of the 
signal must be proportional to the weight squared of the particular 
state.  In a $N+1$ qbit register, and the worst case when all states 
have roughly the same weight, the strength of a reversible measurement 
is $\sim 2^{-N}$ as is the {\it probability\/} of a particular result 
of a projective measurement.  An efficient quantum algorithm is 
explicitly constructed to have a $|R>$ with much less quantum 
entanglement, however this can remain a difficulty for reversible reads.  
In this regard Grover's algorithm presents no problem since the answer 
dominates the register to be measured.  The Shor scheme requires more 
discussion which will be taken up again below.

The first step is to describe an efficient scheme which reduces the 
ESR response of the signal spin to that coming from the two 
computational $|\sigma>|\overline 0>$ states.  This uses no ancilla 
spins other than the signal spin.  As described above, the signal 
comes only from states of the form $|\sigma>|0A>$.  After a first 
gradient pulse, are $\pi$ pulses at the frequencies $\omega^{10}_{n}; 
\, n=2,N$.  A second less selective $\pi$ pulses sequence, which only 
``looks left'', has frequencies $\omega^{0X}_{n}; \, n=N,2$ where $X = 
0\, {\rm or}\, 1$.  (It is implied that pulses are applied at, or 
cover, {\it both\/} $\omega^{00}_{n}$ and $\omega^{10}_{n}$.)  Last is 
a $\pi$ pulse at $\omega^{1}_{1}$.  For the computational states the 
net effect is $|1>|0\ldots 00> \to |0>|1\ldots 10>$ and $|0>|0\ldots 
00> \to |1>|1\ldots 10>$.  If the last spin was reversed this would be 
a complete time conjugation, however this is not necessary since both 
states de-phase in the same manner for a given molecule and so a 
second identical gradient pulse {\it does\/} re-focus the spins and 
the ESR of the signal spin is recovered.  For the non-computational 
states there are two cases.  If the first ``1'' occurs on the third 
spin, the net effect is $|1>|01A> \to |1>|00A^{\prime\prime}>$ and 
$|0>|01A> \to |1>|10A^{\prime\prime}>$ where $A^{\prime\prime}$ is 
some unique mapping of $A$.  Generally the first ``1'' occurs after 
$M$ zeros $|1>|0\ldots 001A> \to |0>|1\ldots 100A^{\prime\prime}>$ 
while $|0>|0\ldots 001A> \to |1>|1\ldots 110A^{\prime\prime}>$ where 
$\ldots$ replaces a series of digits ``1''.  In all cases the resulting 
states differ at the qbit positioned two places before the ``A'' and 
as a result the final states dephase differently during the second 
gradient pulse and the ESR signal is not recovered.  In addition, for 
the first case, the signal spin is not properly conjugated.

This preparation stage permits the extraction of the result of the 
quantum computation on the pure states $|\sigma>U|\overline 0>$.  The 
transformation $U$ does not change the signal spin ESR. The weight of 
any given state $|\sigma_{2}\ldots \sigma_{N}>$ in $U|\overline 0>$ 
might be determined by another gradient-$\pi$-gradient sequence, however 
it would take an exponentially large time $\sim 2^{N}$ to measure the 
amplitude of every state.  There are probably as many strategies for 
rendering practical such a scheme as there are quantum algorithms (so 
finding them is not a big task as of writing).  

The Shor routine searches for the period $r$ of $f(a) = x^{a} \ {\rm 
Mod} N$ where $N$ is a number to be factored and $x<N$ is randomly 
chosen.  The result is $|R> = (1/ \sqrt{r})\sum_{p=0}^{r-1}|pT>|F_p>$, 
and the sought for $r = w/ T$ where $w$ corresponds to the size of the 
registers and where $T$ is the period, i.e., the smallest non-trivial 
number in the first register.  The (normalized) $|F_p>$ need not be 
read.  This smallest number has the largest quantity, $N$, of leading 
zeros, i.e., it is of the form $|\sigma>|0\ldots 0A> \equiv 
|s_{1}=\sigma>|S_{\beta} = 0\ldots 0>_{\alpha}|A>_{\beta}$.  The 
pseudo-code which conjugates the signal spin for such a state is

{\tt If $S_{\beta}=0$, $s_{1} = s_{1} + 1$ (mod 2)}
 
This program line is run with $N=1,2\ldots$ until it fails, with 
$N=N_{0}$, to conjugate the signal spin.  The last such pulse sequence 
is then run backwards to recover the result before the failure.  The 
smallest number then begins $|\sigma>|0\ldots 01A^{\prime}>$ with 
$N_{0}-1$ zeros and the search is continued for the smallest 
$A^{\prime}$, etc.

Again a more efficient equivalent process is possible.  The 
$|\sigma>|0A>$ will exist if a significant signal at $\omega^{0}_{1}$ is 
observed.  Assuming this is the case, a finite amplitude for the 
states $|\sigma>|00A>$ is sought.  First the signal from the $|1A>$ 
states is ``killed'' by a $\pi/2$ pulse at $\omega^{1}_{1}$.  The 
remaining task is relatively simple since it is known where the first 
``error'' will occur.  As above the process starts with a gradient 
pulse, followed by $\pi$ pulses at $\omega^{11}_{2}$ and 
$\omega^{0}_{1}$.  The net changes are $|1>|01A> \to |1>|11A>$, 
$|1>|00A> \to |0>|00A>$, $|0>|00A>\to |1>|00A> $ and $|0>|01A>\to 
|1>|01A>$, i.e., this performs the necessary conjugation on the signal 
spin.  The final gradient pulse will restore the ESR signal from the 
states $|\sigma>|00A>$ while leaving the states originating from 
$|\sigma>|01A>$ with the signal turned off.

The generalization to $|\sigma>|0\ldots 0A>$, where there are $N$ 
zeros, contains the $\pi$ sequences $\omega^{10}_{n}$, $n=2,N-1$; 
$\omega^{11}_{N}$ followed by $\omega^{10}_{n}$, $n=N-1,2$ which 
resets the state $|1>|0\ldots 00A>$ while having the effect 
$|1>|0\ldots 01A> \to |1>|11\ldots 11A>$ but leaving the states 
$|0>|0\ldots 00A>$ and $|0>|00\ldots 01A>$ unchanged.  Clearly the 
sandwiching gradient pulses destroy all but the signal from 
$|\sigma>|0\ldots 0A>$.  The generalization to the search for other 
specific states is evident.

The signal corresponding to $(1/ \sqrt{r})|T>$ is reduced by a factor 
$\sim 1/r$ where $r<N$, $N$ being the number to be factorised.  It is 
possible to detect $\sim 10^{5}$ equivalent spins, so if $r > 10^{18}$ 
even for a $10^{23}$ element ensemble the signal for a $\sim 100$ qbit 
register will become unreadable.  However a pseudo-collapse of the 
wavefunction can be accomplished by changing, a little, Shor's 
algorithm.  The usual calculation, $U$, causes $|a>|b> \to |a>|x^a 
b>$, where again $x$ is a randomly chosen $c$-number less than $N$.  
The pseudo-collapse is accomplished by modifying the second register so 
that the effect of $U$ is $|a>[|1>+|x> + |x^2> + \ldots+ |x^n>+ \ldots 
+ |x^s>] \to |a>[|x^a>+|x^{a+1}> + |x^{a+2}> + \ldots+ |x^{a+n}>+ 
\ldots + |x^{a+s}> ]$ where $s<r$.  This modification can be performed 
efficiently but requires an extra register.  Following an initial 
failure to read $|T>$ with $s=0$, a second calculation with $s \sim 
10^{18}$ must succeed.

For the Grover algorithm the answer dominates $U|\overline 0>$.  First 
the signal spin is observed.  There will be a single strong 
line which determines the leading digit in $U|\overline 0>$.  Imagine, 
e.g., that this implies $U|\overline 0> \sim |1A>$, using the method of 
the previous few paragraphs the amplitude of $|11A>$ is determined.  If, 
e.g., this is small then the state begins $|10A>$ and the last 
interrogation sequence is run backwards to recover the amplitude for 
this state and then $|101A>$ sought for, etc.  (An alternative but 
somewhat more ``rough and ready'' method is to apply a small angle, say 
$\pi/20$, pulse to all spins.  This will turn ``on'' an ESR signal which 
corresponds to the desired state, e.g., if $U|\overline 0> \sim 
|10100010\ldots>$ then the $n=2$ spin will give a free induction 
signal at the two positions $\omega^{X1}_{2}$ and then at 
$\omega^{11}_{3}$, $\omega^{00}_{4}$, $\omega^{10}_{5}$, $\ldots$.)

Error correction represents an interesting challenge.  In order to 
remove the entropy generated by errors, such schemes (17) use 
projective measurements in an essential fashion.  Replacing these by 
the pseudo-projective equivalents would cause the signal strength to 
fall off exponentially.  However it is not hard to find alternative 
means by which to perform the correction using conditional unitary 
transformations.  The entropy must be carried away by ancillary qbits 
and then removed by cooling.  A full discussion of these possibilities 
is too lengthy to be reproduced here.

The author thanks E. Knill for helpful comments.

\vskip 25pt

\vfill\eject

\centerline{\bf REFERENCES AND NOTES}
\vskip 15pt

\item{1.} P. Shor, In {\it Proceedings of the 35th Annual Symposium on 
Foundations of Computer Science}, (IEEE Computer Society, Los 
Alamitos, CA, 1994).

\item{2.} J. I. Cirac  and P. Zoller,  
{\it Phys. Rev. Lett.} {\bf 74} 4091-4094 (1995).

\item{3.} D. Beckman,  A. Chari, S.  Devabhaktuni  and J. Preskill, 
{\it Phys. Rev. A} {\bf 54}, 1034-1063 (1996).

\item{4.} N. A. Gershenfeld  and I. L. Chuang, 
{\it Science} {\bf 275} 350-356 (1997).

\item{5.} D. G. Cory, A. F. Fahmy and T. F. Havel {\it Proc.  of the 
4th Workshop on Physics and Computation}, (New England Complex Systems 
Institute, Boston, MA 1996)

\item{6.} W. S. Warren, {\it Science}, {\bf 277} 1688 (1997).

\item{7.} R. Jones, J. A. Howard, H. A. Joly, P. P. Edwards and R. J. 
Singer, {\it Mag. Resonance in Chem.}, {\bf 33} S98 (1995).

\item{8.} See e.g., J. Veciana, J. Cirujeda, C. Rovira and J. 
Vidal-Gancedo, {\it Adv. Mat.} {\bf 7}, 221 (1995).

\item{9.} D. P. DiVincenzo,
{\it Phys. Rev.} A {\bf 51} 1015-1022 (1995).

\item{10.} see, e.g., O. Kahn, {\it Molecular Magnetism}, (VCH 
Publishers, Inc., 220 East 23rd Street, New York, NY 10010-4606, 
1993), D. Gatteschi, {\it Current Opinion in Solid State and Mat.  
Science}, {\bf 1}, 192 (1996).

\item{11.} S. Lloyd,  {\it 
Science,} {\bf 261} 1569 (1993); see also {\it Science}, {\bf 263} 695 (1994).

\item{12.}  D. G. Cory, M. D. Price and T. F. Havel, quant-phys/970900, (1997).

\item{13.} E. Knill, I Chuang and R. Laflamme, quant-phys/19706053, (1997).

\item{14.} W. Nagourney et al., {\it Phys. Rev. Lett.} {\bf  56}, 2797 (1986); 
J. C. Bergquist et al.  {\it Phys.  Rev.  Lett.} {\bf 56}, 1699 (1986); T. 
Sauter et al., {\it Phys.  Rev.  Lett.} {\bf 56}, 1696 (1986).

\item{15.} See e.g., C. P. Slichter, {\it Principles of 
Magnetic Resonance}, (3rd. Ed., Springer-Verlag, Heidelberg Germany, 1989).

\item{16.} L. K. Grover, {\it Phys. Rev. Lett.}, {\bf 79}, 325 (1997).

\item{17.} See, e.g., D. P. DiVincenzo and P. W. Shor, {\it Phys.  
Rev.  Lett.} {\bf 77}, 3260 (1996) and references therein.

Press, 

\bye